\documentclass[secnumarabic,pre,aps,showpacs,showkeys,twocolumn]{revtex4}
\usepackage{amsmath,amssymb,graphicx}

\newcommand{\np}{\newpage\noindent}

\newcommand{\va}{\mbox{\boldmath $a$}}
\newcommand{\vA}{\mbox{\boldmath $A$}}
\newcommand{\vb}{\mbox{\boldmath $b$}}
\newcommand{\vha}{\hat{\va}{}}
\newcommand{\vhb}{\hat{\vb}{}}
\newcommand{\vB}{\mbox{\boldmath $B$}}

\newcommand{\vj}{\mbox{\boldmath $j$}}
\newcommand{\vdj}{\mbox{\boldmath $\jmath$}}
\newcommand{\vhj}{\hat{\vdj}{}}

\newcommand{\vJ}{\mbox{\boldmath $J$}}

\newcommand{\vvr}{\mbox{\boldmath $r$}}

\newcommand{\vu}{\mbox{\boldmath $u$}}
\newcommand{\vv}{\mbox{\boldmath $v$}}
\newcommand{\vV}{\mbox{\boldmath $V$}}

\newcommand{\hj}{\hat{\jmath}{}}
\newcommand{\balf}{\mbox{\boldmath $\alpha$}}
\newcommand{\bbet}{\mbox{\boldmath $\beta$}}
\newcommand{\vsig}{\mbox{\boldmath $\sigma$}}
\newcommand{\vrho}{\mbox{\boldmath $\rho$}}
\newcommand{\OC}{\mbox{\boldmath $\rm C$}}
\newcommand{\OP}{\mbox{\boldmath $\rm P$}}
\newcommand{\OQ}{\mbox{\boldmath $\rm Q$}}
\newcommand{\OS}{\mbox{\boldmath $\rm S$}}
\newcommand{\OT}{\mbox{\boldmath $\rm T$}}

\newcommand{\tauc}{\tau_{\rm c}}
\newcommand{\lc}{\lambda_{\rm c}}

\newcommand{\mx}{\langle x\rangle}
\newcommand{\mvB}{\langle\vB\rangle}

\newcommand{\na}{\nabla}
\newcommand{\pa}{\partial}
\newcommand{\md}{{\rm d}}
\newcommand{\me}{{\rm e}}
\newcommand{\hlf}{{\textstyle\frac{1}{2}}}

\begin{document}

\title{\bf Statistical dynamo theory: Mode excitation}

\author{P. Hoyng}
\email{p.hoyng@sron.nl}
\affiliation{SRON Netherlands Institute for Space Research, \\
Sorbonnelaan 2, 3584 CA Utrecht, The Netherlands}

\date{\today}

%%%%%%%%%%%%%%%%%%%%%%%%%%%%

\begin{abstract}
We compute statistical properties of the lowest-order multipole coefficients
of the magnetic field generated by a dynamo of arbitrary shape. To this end
we expand the field in a complete biorthogonal set of base functions, viz.
$\vB=\sum_k a^k(t)\vb^k(\vvr)$. The properties of these biorthogonal function
sets are treated in detail. We consider a linear problem and the statistical
properties of the fluid flow are supposed to be given. The turbulent
convection may have an arbitrary distribution of spatial scales. The time
evolution of the expansion coefficients $a^k$ is governed by a stochastic
differential equation from which we infer their averages $\langle a^k\rangle$,
autocorrelation functions $\langle a^k(t)a^{k*}(t+\tau)\rangle$, and an
equation for the cross correlations $\langle a^ka^{\ell*}\rangle$. The
eigenfunctions of the dynamo equation (with eigenvalues $\lambda_k$) turn out
to be a preferred set in terms of which our results assume their simplest
form. The magnetic field of the dynamo is shown to consist of transiently
excited eigenmodes whose frequency and coherence time is given by
$\Im\lambda_k$ and $-1/\Re\lambda_k$, respectively. The relative r.m.s.
excitation level of the eigenmodes, and hence the distribution of magnetic
energy over spatial scales, is determined by linear theory. An expression is
derived for $\langle\vert a^k\vert^2\rangle/\langle\vert a^0\vert^2\rangle$
in case the fundamental mode $\vb^0$ has a dominant amplitude, and we outline
how this expression may be evaluated. It is estimated that $\langle\vert a^k
\vert^2\rangle/\langle\vert a^0\vert^2\rangle\sim 1/N$ where $N$ is the
number of convective cells in the dynamo. We show that the old problem of a
short correlation time (or FOSA) has been partially eliminated. Finally we
prove that for a simple statistically steady dynamo with finite resistivity
all eigenvalues obey $\Re\lambda_k<0$.
\end{abstract}

%%%%%%%%%%%%%%%%%%%%%%%%%%%%

\maketitle

%%%%%%%%%%%%%%%%%%%%%%%%%%%%

\section{Introduction}
\label{sec:intro}
The origin of the magnetic field of the Earth and the Sun is well understood
at the qualitative level. Helical convection acting on the toroidal component
of the field generates new poloidal field, while the toroidal field is
regenerated either by shear flows acting on the poloidal field or by the same
helical convection now acting on the poloidal field. These dynamo processes
are in principle able to balance resistive decay and to maintain a magnetic
field for very long times. The large scale magnetic fields observed in 
galactic discs are likewise believed to be due to similar dynamo processes 
\cite{RH04}. Self-consistent hydromagnetic simulations that became available 
since 1995 have confirmed this dynamo picture for the geomagnetic field 
\cite{GR95}. Many groups have since then published numerical geodynamo models. 
Even though the available computational means do not permit the parameters of 
the models to be `earth-like', the magnetic field of the simulations has many 
properties in common with the observed geomagnetic field \cite{GR95,KB97,%
COG98,TMH05,CW07}. For the Sun with its much higher magnetic Reynolds number 
such simulations are not yet feasible. 

The availability of numerical geodynamo models has opened up the possibility
for a detailed diagnostics of dynamo action. Kageyama and Sato \cite{KS97} and
Olson et al. \cite{OCG99} found that the regeneration of poloidal and toroidal
field resembles the $\alpha^2$-dynamo scenario of mean field theory 
\cite{KR80}. Wicht and Olson \cite{WO04} analysed the sequence of events
leading to a reversal in a simple numerical dynamo model. Schrinner et al.
\cite{SRSRC07} have inferred mean field tensors $\alpha_{ik}$ and
$\beta_{ik\ell}$ and the mean field $\mvB$ from simulations. The mean field is
then compared with the mean field predicted by the dynamo equation. Reasonable
agreement was found for a simple magnetoconvection model and for simple dynamo
models. For a recent review we refer to Wicht et al. \cite{WSH08}.

The evolution of the magnetic field $\vB$ in the conducting fluid of a dynamo
is governed by the induction equation:
\begin{equation}
\frac{\pa\vB}{\pa t}\,=\,\na\times(\vV\,-\,\eta\na\,)\times\vB\ .
\label{eq:induct}
\end{equation}
The dynamo is located in a volume $V$ with exterior vacuum $E$, see
Fig.~\ref{fig:dynmodel}. The shape of the dynamo need not be spherical. The
flow $\vV$ consists of a stationary component $\vv$ and the turbulent
convection $\vu$, which may have an arbitrary distribution of spatial scales:
\begin{equation}
\vV\,=\,\vv+\vu\ .
\label{eq:vV}
\end{equation}
A popular line of attack is to average over the turbulent convection $\vu$
and to derive an equation for the mean field $\mvB$ (the so-called mean field
dynamo equation or briefly dynamo equation \cite{M78,KR80}). We follow a
different path, and we expand the field $\vB$ in a complete set of functions
$\vb^i(\vvr)$. We then determine the statistical properties of the expansion
coefficients. The mean field $\mvB$ will appear only occasionally, as a
mathematical concept without much physical meaning attached to it.

The idea to study dynamo-generated magnetic fields by an expansion in
multipoles goes back to Elsasser \cite{E46}. We follow the same technique and
obtain a set of equations for the expansion or mode coefficients. These
equations contain a random element as the fluid motion $\vV$ in the induction
equation consists of a steady part with a superposed turbulent convective
component. The new aspect is that we use the theory of stochastic differential
equations \cite{vK76} to infer the statistical properties of the mode
coefficients. We consider a statistically steady, saturated dynamo with a
selfconsistent mean flow $\vu$ and turbulent flow $\vv$.

We treat a linear problem and consider $\vu$ and $\vv$ as given. Until
recently it had been tacidly assumed that the solution of the induction
equation (\ref{eq:induct}) represents the selfconsistent field $\vB$ obtained
from a nonlinear solution of the MHD equation (provided one uses the exact
selfconsistent flow $\vu+\vv$). But we now know that this is not correct.
There are statistically steady, saturated dynamos whose velocity field, taken
as a given input flow in the induction equation, acts as a kinematic dynamo
with exponentially growing solutions \cite{CT09,TB08}. In those cases the
induction equation on its own is obviously unable to reproduce the
selfconsistent field $\vB$. Many questions regarding this unexpected
phenomenon remain to be answered. For example, it is not known to what extent 
it is a universal feature. Schrinner and coworkers (to be submitted) have 
found several counterexamples, and their results suggest that the flow fields 
of fast rotating geodynamo models are also kinematically stable. The solution 
of the induction equation with the (selfconsistent) flow taken from these 
dynamos is (after a transitory period) up to a constant factor equal to the
selfconsistent field $\vB$, independent of the initial condition. In the
absence of a generally agreed-upon terminology we shall in this paper refer
to these dynamos as {\em kinematically stable dynamos}.

Here we restrict ourselves to kinematically stable dynamos, so that the
solution of (\ref{eq:induct}) faithfully represents the actual field $\vB$.
Otherwise, the dynamo model is general and may be of the geodynamo or solar
type. We take the existence of a linear dynamo instability and its nonlinear
saturation for granted, and study the fluctuations of the system around this
nonlinear equilibrium state driven by the turbulent convection. A proviso must 
be made for galactic dynamos since these may not yet be in a statistically 
steady saturated state. It is therefore not clear if the theory developed here 
is applicable to these dynamos.   

In this study we do not focus on the mean field concept $\mvB$; we rather
expand the field $\vB$ in a complete set of functions $\vb^i(\vvr)$. But
as we determine the statistical properties of the expansion coefficients,
notions from mean field theory pop up simply because we compute averages.
For example, the dynamo parameters $\alpha$ and $\beta$, well known from
mean field theory, emerge because they are connected to the simplest
nontrivial averages of the turbulent convection. It is almost unavoidable
that they should appear in any theory that considers averages over the
turbulent convection. In other words, as we compute the statistical
properties of the expansion coefficients we make contact with mean field
{\it theory}, but we do not use the mean field {\it concept} $\mvB$, except
occasionally in passing.

The context of this study is as follows. Hoyng et al. \cite{HOS01} analysed
the statistical properties of the multipole coefficients of the mean field of
a simple mean field geodynamo model excited by fluctuations in the dynamo
parameter $\alpha$. Here we addess a much more general problem: the
statistical properties of the multipole coefficients of the field itself due
to forcing by the turbulent convection. The present paper builds on an
earlier study of Hoyng \cite{H88} that addressed the same problems, but with
limited success. The question of the r.m.s. mode excitation level was only
solved conceptually, and the derivation of the time evolution of the mean
expansion coefficients contained errors. Moreover, the study was restricted
to dynamos with homogeneous isotropic turbulence. These points of criticism
have been removed here. Since we study the relative excitation level of
magnetic overtones in the dynamo, the present work may also be regarded as a
spectral theory \cite{PFL76} yielding the distribution of magnetic energy
in a finite dynamo as a function of spatial scale (i.e. mode number), with 
due allowance for the boundary conditions.

This paper is organised as follows. In Sec.~\ref{sec:compset} we summarize
the properties of the function sets we use to represent the magnetic field of
the dynamo, in particular biorthogonality and closure relations. Then we
derive the equation for the mode expansion coefficients in
Sec.~\ref{sec:modeq}. Next we determine the statistical properties of the
expansion coefficients. In Sec.~\ref{sec:avmod} we derive the time evolution
of the mean mode coefficients and show that it is governed by the operator
of the dynamo equation. In Sec.~\ref{sec:rmsmod} and \ref{sec:omag} we compute
the excitation level of the overtones relative to the fundamental (dipole)
mode. Finally, in Sec.~\ref{sec:disc} we discuss our results and potential
applications in a somewhat wider perspective.

%%%%%%%%%%%%%%%%%%%%%%%%%%%

\section{Complete function sets}
\label{sec:compset}
We summarize here some technicalities concerning the function sets that we
employ. Some of these properties have been derived elsewhere \cite{H88,HS95},
and we collect them here in the interest of a homogeneous notation, taking
the opportunity to correct some mistakes. Readers may move directly to the
next Section, and return here for reference.

\begin{figure}
%\vspace{4.cm}
\centerline{\includegraphics[width=4.cm]{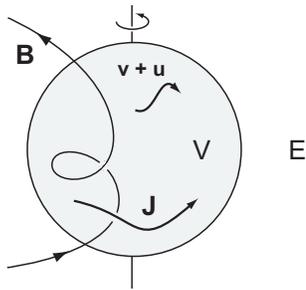}}
\caption{\small A dynamo in a volume $V$ that need not be spherical, with
exterior vacuum $E$. The fluid in $V$ has a magnetic resistivity $\eta$, and
the flow $\vV$ consists of a stationary component $\vv$ and the turbulent
convection $\vu$, which may have an arbitrary distribution over spatial
scales. The currents $\vJ$ are restricted to $V$, and both currents and flows
are tangential to the boundary $\pa V$. The magnetic field $\vB$ is continuous
through $\pa V$.
\vspace{5.mm}}
\label{fig:dynmodel}
\end{figure}

We expand the field $\vB$ of the dynamo, see Fig.~\ref{fig:dynmodel}, in a
complete set of functions $\vb^j(\vvr)\,$:
\begin{equation}
\vB\,=\,\sum_j a^j(t)\,\vb^j(\vvr)\ .
\label{eq:exp}
\end{equation}
These functions are often eigenfunctions of some differential operator,
and since this operator is usually not self-adjoint the functions are not
orthogonal. This problem is handled by using the adjoint set $\vhb^i(\vvr)$,
with the following property
\begin{equation}
\int_{V+E}\,\vhb^i\cdot\vb^j\,\md^3\vvr\,=\,\delta^{ij}\ .
\label{eq:biorth}
\end{equation}
The base functions $\vhb^i(\vvr)$ and $\vb^j(\vvr)$ constitute a biorthonormal
set \cite{MF53,H88}; they are continuous through the boundary $\pa V$ of $V$, 
just like $\vB$, and potential in $E$. A few remarks on the notation:
\\[2.mm]
- the hat $\hat{{\ }}$ includes complex conjugation, so
\[\hat{\vB}\,=\,\sum_ia^{i*}(t)\,\vhb^i(\vvr)\,,\]
- upper indices enumerate the base functions, \\
- lower indices indicate vector components, \\
- currents are defined as $\vJ=\na\times\vB$ -- the factor $4\pi/c$ is
absorbed in $\vJ$.
\\[2.mm]
%
%%%%%%%%%%%%%%%%%%%%%%%%%%%
%

%%%%%%%%%%%%%%%%%%%%%%%%%%%%

Examples of such function sets are the free decay modes, or the eigenfunctions
of a homogeneous spherical dynamo (see Ref.~\cite{KR80}, Chap. 14). These two
sets are actually self-adjoint so that the adjoint sets are obtained by
complex conjugation: $\vhb^i(\vvr)=\vb^{i*}(\vvr)$. Usually the situation is
more complicated, and we refer to Schrinner et al. \cite{SJSH08} for the
construction of the adjoint set of the eigenfunctions of the dynamo equation.
From that construction a number of properties emerge, such as $\vhj\equiv$
adjoint of $\na\times\vb\equiv\na\times\vhb$, i.e. the adjoint operation
commutes with $\na$.

%%%%%%%%%%%%%%%%%%%%%%%%%%%%

\subsection{Biorthogonal sets in $V$}
\label{sec:biorthV}
It is of course possible to define an inner product in $V$ alone, and to
construct an adjoint set $\vhb^i$ of $\vb^j$ so that these are biorthogonal in
$V$: $\int_V \vhb^i\cdot\vb^j\,\md^3\vvr\,=\,\delta^{ij}$. But as we shall see
these sets are not very convenient. Fortunately, the boundary conditions
permit construction of a very useful biorthogonal set in $V$, starting from
sets $\{\vhb^i\}$ and $\{\vb^j\}$ that are biorthogonal in $V+E$. For two
fields $\vb_1$ and $\vb_2$ with associated currents $\vj=\na\times\vb$ and
vector potentials $\va$, where $\na\times\va=\vb$, we have $\na\cdot\vhb_1
\times\va_2=\vhj_1\cdot\va_2-\vhb_1\cdot\vb_2$, whence
\begin{eqnarray}
\int_{V+E}\vhb_1\cdot\vb_2\,\md^3\vvr&=&
\nonumber \\[1.mm]
\int_V\vhj_1\cdot\va_2\,\md^3\vvr&-&
\left(\oint_{\pa E}+\oint_{\pa V}\right)
\vhb_1\times\va_2\cdot\md^2\vsig\ .\quad
\label{eq:biorth1}
\end{eqnarray}
Here $\pa V$ and $\pa E$ indicate the boundary of $V$ and $E$, respectively.
The surface integrals cancel because the field is continuous through $\pa V$
and the vector potential can be constructed to be so. Volume integrals
containing currents may be restricted to $V$ since $\vj=0$ in $E$. This leads
to:
\begin{eqnarray}
\int_{V+E}\vhb^i\cdot\vb^j\,\md^3\vvr&=&
\int_V\vhj^i\cdot\va^j\,\md^3\vvr
\nonumber \\[2.mm]
&=&\int_V\vha^i\cdot\vj^j\,\md^3\vvr\,=\delta^{ij}\ .
\label{eq:biorth2}
\end{eqnarray}
These relations are invariant under a gauge transformation $\va\rightarrow\va+
\na\psi$ because $\int_V\vhj\cdot\na\psi\,\md^3\vvr=\int_V\na\cdot(\psi\vhj)\,
\md^3\vvr=\int_{\pa V}\psi\,\vhj\cdot\md^2\vsig=0$, as (adjoint) currents run
parallel to the boundary. Currents and vector potentials thus form a
biorthogonal set in $V$, and this is essential for our study.

%%%%%%%%%%%%%%%%%%%%%%%%%%%%%%%%%

\subsection{Closure relations}
\label{sec:closure}
Completeness means that any reasonable field $\vB$ should have a unique
expansion (\ref{eq:exp}). Completeness implies the existence of closure
relations. The best known example is from quantum mechanics where
orthonormality of two eigenstates $m$ and $n$ means that $\int\psi_m(\vvr)^*
\psi_n(\vvr)\,\md^3\vvr=\delta_{mn}$, and the closure relation is $\sum_n
\psi_n(\vvr)^*\psi_n(\vvr')=\delta(\vvr-\vvr')$, provided the sum is over a
complete set of eigenstates \cite{D78}. In the present context it can be shown
that
\begin{eqnarray}
\sum_i\,\hat{b}_{k2}^i\,b_{\ell 1}^i\,=
\sum_i\,\hj_{k2}^i\,a_{\ell 1}^i&=&
\sum_i\,\hat{a}_{k2}^i\,j_{\ell 1}^i
\nonumber \\[1.mm]
&=&\delta({\it 1}-{\it 2})\,\delta_{k\ell}\ .
\label{eq:complet2}
\end{eqnarray}
Notation: $\it 1=\vvr_1, \it 2=\vvr_2$, $b_{\ell 1}^i=\ell$-th vector component
of $\vb^i(\vvr_1)$, etc. Relation~(\ref{eq:complet2}) says that the expressions
on the left are unit operators in the discrete and continuous indices. There
are several other closure relations, and the one we actually need is
\begin{equation}
\sum_i\,\hj_{k2}^i\,b_{\ell 1}^i\,=\,\epsilon_{k\ell m}\,
\delta({\it 1}-{\it 2})\na_{m2}\ .
\label{eq:complet3}
\end{equation}
Relations (\ref{eq:complet2}) and (\ref{eq:complet3}) hold under an integral
sign $\int_V\md{\it 2}\cdots$ and are to be regarded as an operator acting on
all $\vvr_2$ dependence in the integrand. Care is needed when applying these
closure relations, and for details we refer the reader to Appendix C of Ref.
\cite{HS95}. Recall that we have absorbed the factor $4\pi/c$ in the
definition of the current.

%%%%%%%%%%%%%%%%%%%%%%%%%%%%%%%%%%%%%%%

\section{Mode equations}
\label{sec:modeq}
On taking the inner product of (\ref{eq:exp}) with $\vhb^k$ we obtain with the
help of (\ref{eq:biorth}):
\begin{equation}
a^k(t)\,=\,\int_{V+E}\,\vhb^k\cdot\vB\;\md^3\vvr\ .
\label{eq:akt1}
\end{equation}
To obtain the evolution equation of $a^k$, we take the time derivative 
$\dot{{\ }}=\pa_t\/$ and split the integral in two parts:
\begin{equation}
\dot{a}^k(t)\,=\,\int_V\,\vhb^k\cdot\dot{\vB}\;\md^3\vvr\,+\,
                 \int_E\,\vhb^k\cdot\dot{\vB}\;\md^3\vvr\ .
\label{eq:dota1}
\end{equation}
In the first term we may use the induction equation, but we have no equation
for $\pa_t\vB$ in $E$, where $\vB$ is simply the potential field continuation
of $\vB$ on $\pa V$. Omitting the second term is no option as it amounts to
saying that $\pa_t\vB=0$ in $E$. A way out of this dilemma is to integrate 
the right hand side of (\ref{eq:akt1}) by parts, as in (\ref{eq:biorth1}), to
obtain ($\na\times\vA=\vB$):
\begin{equation}
a^k(t)\,=\,\int_V\,\vhj^k\cdot\vA\;\md^3\vvr\ .
\label{eq:akt2}
\end{equation}
Derivation of the mode equations is now straightforward. Take the time
derivative of (\ref{eq:akt2}) and insert the uncurled induction equation,
$\pa_t\vA=\{(\vv+\vu)-\,\eta\na\}\times\vB+\na\psi$, to obtain
\begin{equation}
\dot{a}^k(t)\,=\,\int_V\,\vhj^k\cdot\{(\vv+\vu)-\,\eta\na\}\times\vB\;
\md^3\vvr\ .
\label{eq:dota2}
\end{equation}
The gradient term produces zero, see below (\ref{eq:biorth2}). At this point
we note that if we had used a biorthogonal set $\vb^i(\vvr),\,\vhb^j(\vvr)$ in
$V$, the dilemma mentioned a moment ago would not occur, because the integrals
are restricted to $V$. However (\ref{eq:dota2}) would then read
\begin{equation}
\dot{a}^k(t)\,=\,\int_V\,\vhb^k\cdot\na\times\{(\vv+\vu)-\,\eta\na\}
\times\vB\;\md^3\vvr\ .
\label{eq:dota3}
\end{equation}
There is no conflict between (\ref{eq:dota2}) and (\ref{eq:dota3}) because 
they use different bases: biorthogonal in $V+E$ and biorthogonal in $V$, 
respectively. So $\vhb^k,\,\vhj^k$ in (\ref{eq:dota2}) are not the same as 
$\vhb^k,\,\vhj^k$ in (\ref{eq:dota3})! 

The extra $\na$ on the right hand side of (\ref{eq:dota3}) is very 
inconvenient and cannot be removed by integration by parts because the surface 
integral over $\pa V$ is nonzero (see Appendix~\ref{sec:formmod}). It would 
stay with us and cause havoc in all subsequent analytical and numerical work. 
Therefore (\ref{eq:dota2}) is much to be preferred. This point was one of the 
main drivers for developing the formalism of the previous Section. Because of 
its importance, we discuss the issue in more detail in 
Appendix~\ref{sec:formmod}.

On inserting $\vB=\sum_\ell a^\ell\,\vb^\ell$ in (\ref{eq:dota2}), we arrive at
the mode equations
\begin{eqnarray}
\dot{a}^k&=&\sum_\ell\,\{S^{k\ell}+C^{k\ell}(t)\}\,a^\ell\ ,
\label{eq:modeq1} \\[0.mm]
\lefteqn{\rm with}
\nonumber \hspace{2.cm} \\[0.mm]
S^{k\ell}&=&\int_V\;\vhj^k\cdot(\vv-\,\eta\na\,)\times\vb^\ell\;
\md^3\vvr\ ;\qquad\qquad
\label{eq:rkl} \\[2.mm]
C^{k\ell}(t)&=&\int_V\;\vhj^k\cdot\vu\times
\vb^\ell\;\md^3\vvr
\nonumber \\[-2.mm]
\label{eq:ckl} \\[-2.mm]
&=&-\int_V\;\vu\cdot\vhj^k\times
\vb^\ell\;\md^3\vvr\ .
\nonumber
\end{eqnarray}
The base functions that occur in the definition of the matrix elements
$S^{k\ell}$ and $C^{k\ell}$ may be thought of as spatial filters with positive 
and negative signs indicating how the velocities $\vv$ and $\vu$ are to be 
weighted in the volume integration over $V$. The boundary conditions make that
the integrals are of the type $\langle{\rm current}|{\rm operator}|{\rm field}
\rangle$ rather than $\langle{\rm field}|{\rm operator}|{\rm field}\rangle$.
The boundary conditions have therefore a profound effect on the definition of
these filters. Mode equations of the type (\ref{eq:modeq1}) had already been
derived by Elsasser \cite{E46}, although he considered steady flows only.
Compared to the time scales in the steady matrix $S^{k\ell}$, the elements
$C^{k\ell}$ fluctuate rapidly because $\vu(\vvr,t)$ does. The mode equations
are therefore stochastic equations, and we shall use the theory of these
equations to infer averages of the mode coefficients.

%%%%%%%%%%%%%%%%%%%%%%%%%%%%%%%%%%%%%%%%

\section{Stochastic equations with multiplicative noise}
\label{sec:stocheq}
Consider a linear equation of the type
\begin{equation}
\frac{\pa x}{\pa t}\,=\,\{A+F(t)\}x\ ,
\label{eq:stocheq1}
\end{equation}
where linear means linear in $x$. The quantity $x$ may be anything, a scalar,
a vector such as $\vB$, or other. The operator $A$ is independent of time but
$F(t)$ fluctuates randomly, with a correlation time $\tauc$ and zero mean,
$\langle F\rangle\,=\,0$. This `noise term' $F(t)$ is said to be 
multiplicative as it multiplies the independent variable $x$, and that makes
(\ref{eq:stocheq1}) more complicated than for example $\pa x/\pa t\,=\,Ax+
F(t)$, where the noise term is additive.

A tenet of the theory of stochastic differential equations is that the average
$\langle x\rangle$ obeys a closed equation \cite{vK76}:
\begin{eqnarray}
\frac{\pa}{\pa t}\,\mx&=&\biggl\{A\,+
\int_0^\infty\md\tau\,\langle F(t)\exp(A\tau)\cdot\qquad\qquad
\nonumber \\
&&\qquad\qquad\cdot\,F(t-\tau)\rangle\exp(-A\tau)\biggr\}\mx\ ,
\label{eq:stocheq2}
\end{eqnarray}
The right hand side of (\ref{eq:stocheq2}) is actually the first term of a 
series with terms of order $(F\tauc)^n$ containing higher-than-second-order 
correlation functions \cite{vK74}. Here $F$ stands for $F_{\rm r.m.s.}$. We 
assume that the correlation time is small, $F\tauc\ll 1$, so that only the 
first term survives and only second order correlations matter, as in 
(\ref{eq:stocheq2}). Frequently also $A\tauc\ll 1$ holds, and then we may 
ignore the exponential operators in (\ref{eq:stocheq2}). These evolution 
operators $\exp(\pm A\tau)$, incidentally, ensure that the transport 
coefficients for $\mx$ derived from (\ref{eq:stocheq2}) are Galilean invariant 
\cite{H03}. Relation (\ref{eq:stocheq2}) will be applied several times in what 
follows.

%%%%%%%%%%%%%%%%%%%%%%%%%%%%%%%%%%%%%%%%%%%%%%

\section{The average mode coefficients}
\label{sec:avmod}
The equation for the average $\langle a^k\rangle$ is in principle available in
the literature, but hidden in two papers \cite{H88,HS95}. We summarise the
derivation here, and apply (\ref{eq:stocheq2}) to (\ref{eq:modeq1}):
\begin{equation}
\frac{\md}{\md t}\,\langle a^k\rangle\,=\,\left\{\,S^{k\ell}\,+\,
\int_0^\infty\!\md\tau\,\langle C^{ki}(t)C^{i\ell}(t-\tau)\rangle\,\right\}
\langle a^\ell\rangle\ .
\label{eq:dotma}
\end{equation}
Summation over double upper indices (here $i$ and $\ell$) is implied.
Before we elaborate the correlation function we verify the conditions to be
satisfied:

(1) $C^{k\ell}\tauc\ll 1$, the usual condition of a short correlation time,
now in the context of the present application [\,$C^{k\ell}$ is understood to 
indicate a characteristic magnitude\,]. We estimate the value of $C^{k\ell}\tauc$ 
for the geodynamo. The volume integration in (\ref{eq:ckl}) reduces $\vu$ 
effectively to $u_{\rm r.m.s.}N^{-1/2}$ where $N\sim 4R^3/\lc^3\sim$ number of 
convection cells; $R$ is the radius of the outer core, and $\lc$ the 
correlation length (the volume of the inner core is negligible). We estimate 
$\vhj^k=\na\times\vhb^k\sim\hat{b}^k/(R/k)$ and $\int_V\hat{b}^k\,b^\ell\md^3
\vvr=O(1)$, and $u_{\rm r.m.s.}\tauc\sim\lc$. This leads to
\begin{eqnarray}
C^{k\ell}\tauc&\sim&u_{\rm r.m.s.}\,\tauc\left(\frac{4R^3}{\lc^3}
\right)^{-1/2}\!\!\int_V\frac{\hat{b}^k\,b^\ell}{R/k}\;\md^3\vvr
\nonumber \\[2.mm]
&\sim&\frac{k}{2}\;\left(\frac{\lc}{R}\right)^{5/2}\ .
\label{eq:ckltau}
\end{eqnarray}
For the geodynamo we have $R/\lc\sim 3.5$. It follows that $C^{k\ell}\tauc\,
\sim\,0.02k$ so that $C^{k\ell}\tauc\ll 1$ seems well satisfied for the
lowest multipole coefficients $a^k$ (even though $u_{\rm r.m.s.}\tauc\sim
\lc\,$!). There is an interesting connection here with the problem of the 
First Order Smoothing Approximation (FOSA) in mean field theory, and we refer 
to Sec.~\ref{sec:disc} where we illustrate how this FOSA problem seems to be 
partially eliminated.

(2) $S^{k\ell}\tauc\ll 1$. The magnitude of $S^{k\ell}$ is set by the
resistive term in (\ref{eq:rkl}), and in much the same way we obtain
\begin{equation}
S^{k\ell}\tauc\,\sim\,k\ell\,\eta\tauc/R^2\ .
\label{eq:rkltau}
\end{equation}
For the geodynamo we have $\eta\sim 1\,$m$^2\,$s$^{-1}$, and $S^{k\ell}\tauc\,
\sim\,3\cdot10^{-4}k\ell$ so that $S^{k\ell}\tauc\ll 1$ holds except for high
order modes. We have to assume therefore that these modes contribute little,
which is plausible because they will have small amplitudes.

The correlation function in (\ref{eq:dotma}) is evaluated in
Appendix~\ref{sec:appa}, and as a result we may write (\ref{eq:dotma}) as
follows:
\begin{equation}
\frac{\md}{\md t}\,\langle a^k\rangle\,=\,D^{k\ell}\langle a^\ell\rangle\ ,
\label{eq:dotmaf}
\end{equation}
with
\begin{equation}
D^{k\ell}\,=\,\int_V\;\vhj^k\cdot D\vb^\ell\;\md^3\vvr\ ,
\label{eq:dkl}
\end{equation}
where the operator $D$ is defined as
\begin{equation}
D\vb\,=\,\vv\times\vb+\balf\cdot\vb-
\bbet:(\na\vb)-\eta\na\times\vb\ .
\label{eq:defd}
\end{equation}
Notation: $(\balf\cdot\vb)_i=\alpha_{ik}b_k$ and $[\bbet:(\na\vb)]_i=
\beta_{ik\ell}(\na_kb_\ell)$. The operator $D$ is intimately related to
the dynamo equation, which reads (Appendix~\ref{sec:appa}):
\begin{equation}
\frac{\pa}{\pa t}\,\mvB\,=\,\na\times D\mvB\ .
\label{eq:dyneq}
\end{equation}
The fact that the operator of the dynamo equation for the mean field emerges
in (\ref{eq:dotmaf}) is quite independent of which set of base functions is
used. At this point we see how mean field theory emerges as we study the
statistical properties of the mean mode coefficients. It happens because we
take averages, and since the simplest averages over the turbulent convection
involve $\vv,\,\balf$ and $\bbet$, it is perhaps not surprising that the
operator of the dynamo equation for the mean field appears. But we use the
mean field $\mvB$ as a mathematical concept only, without much of a physical
meaning attached to it.

%%%%%%%%%%%%%%%%%%%%%%%%%%%%%%%%%%%%%%%%%%%%%%%%%%%%%%%

\subsection{Dynamo action and dynamo equation}
\label{sec:meandyneq}
The fact that the dynamo equation for the mean emerges as we set up the theory
indicates a deep connection between dynamo action in flows with a random 
element and the mean-field dynamo equation. What this relation is can be seen 
by adopting the eigenfunctions of the dynamo equation as the complete function 
set $\{\vb^i\}$, with eigenvalues $\lambda_i$ \cite{lowind}. Then we have
\begin{equation}
\na\times D\vb^i\,=\,\lambda_i\,\vb^i\ .
\label{eq:eigenf1}
\end{equation}
Relations (\ref{eq:eigenf1}) through (\ref{eq:corraa}) below hold only if we
use the eigenfunctions of the dynamo equation as our function set. Schrinner
et al. \cite{SJSH08} show how these eigenfunctions and their adjoints may be
constructed.

The computation of $D^{k\ell}$ from (\ref{eq:dkl}) has several potential
pitfalls. Integration by parts to $\int_V\vhb^k\cdot\na\times D\vb^\ell\,
\md^3\vvr$, with hopes of using (\ref{eq:eigenf1}), is not possible because 
the surface term $\oint_{\pa V}\vhb^k\times D\vb^\ell\cdot\md^2\vsig$ does not
vanish. And to argue that (\ref{eq:dkl}) $=\int_{V+E}\,\vhj^k\cdot D\vb^\ell\,
\md^3\vvr=\int_{V+E}\,\vhb^k\cdot\na\times D\vb^\ell\,\md^3\vvr$ does not work
either because $D$ is not defined in $E$. A safe way is to uncurl
(\ref{eq:eigenf1}) to $D\vb^i\,=\,\lambda_i\,\va^i+\na\phi$, and to insert
that in (\ref{eq:dkl}):
\begin{equation}
D^{k\ell}\,=\,\lambda_\ell\int_V\;\vhj^k\cdot\va^\ell\;\md^3\vvr\,=\,
\lambda_\ell\,\delta^{k\ell}\ .
\label{eq:dkl1}
\end{equation}
The gradient term produces zero, see below (\ref{eq:biorth2}). The detour via
the vector potential in (\ref{eq:dkl1}) allows us to avoid trouble due to the
discontinuity of the operator $D$ on $\pa V$.

Returning to the evolution of the mean mode coefficients, we now have
\begin{equation}
\frac{\md}{\md t}\,\langle a^k\rangle\,=\,\lambda_k\langle a^k\rangle\ ,
\quad{\rm or}\quad
\langle a^k\rangle\,\propto\,\exp(\lambda_kt)\ .
\label{eq:dotmaf1}
\end{equation}
If we choose the eigenfunctions of the dynamo equation (the operator
$\na\times D$) as a basis to represent the magnetic field, then the mean mode
coefficients decay as simple exponentials at a rate set by the eigenvalue of
the mode. Representation of the field $\vB$ of the dynamo in terms of the
eigenfunctions of its dynamo equations is therefore an efficient way to expose
the underlying physics. The autocorrelation of the expansion coefficients
becomes:
\begin{equation}
\langle a^k(t+\tau)a^{k*}(t)\rangle\,=\,\langle\vert a^k\vert^2\rangle
\cdot\exp(\lambda_k\tau)
\label{eq:corraa}
\end{equation}
(no summation over $k$). The proof is simple and not given here. The mean
squares of the expansion coefficients are computed in Sec.~\ref{sec:omag},
relation (\ref{autocorr}).

Of course one may always use a different representation. In that case
$D^{k\ell}$ is no longer diagonal and the mean mode coefficients have a more
complicated time dependence. Equation (\ref{eq:dotmaf}) may be integrated to
\begin{equation}
\langle a^k\rangle\vert_t\,=\,[\exp(Dt)]^{k\ell}\,
\langle a^\ell\rangle\vert_{t=0}\ .
\label{eq:expmix}
\end{equation}
The time dependence of $\langle a^k\rangle\vert_t$ and the autocorrelation
function is now a mixture of exponentials.

%%%%%%%%%%%%%%%%%%%%%%%%%%%%%%%%%%%%%%%%%%%%%%%%%%%%%%%

\subsection{Absence of growing modes}
\label{sec:nogrow}
For the theory developed here to make sense it is necessary that all
eigenvalues have a negative real part, $\Re\lambda_k<0$, and we shall assume
that this is the case. The implication is that if we measure the flow in a
dynamo(model) with all feedbacks of the Lorentz force acting on it, we should
find, on solving (\ref{eq:eigenf1}) with $\alpha_{ij}$ and $\beta_{ijk}$ and
mean flow computed from these measurements, that all eigenvalues have
$\Re\lambda_k<0$. This is one of the checks on a correct determination of the
dynamo coefficients. Of course this can only be true if we restrict ourselves
to kinematically stable dynamos (see Sec.~\ref{sec:intro} for a definition).
A proof that $\Re\lambda_k<0$ has been given for a few special cases with
infinite conductivity \cite{H87,vGH89}, but a general proof is lacking. In
Appendix~\ref{sec:relneg} we push the issue one step forward and prove that
all eigenvalues of the dynamo equation of a statistically steady dynamo with
locally isotropic turbulence and finite conductivity have negative real parts.

A consequence is that all mean mode coefficients approach zero, $\langle a^k
\rangle\downarrow 0$ and that the mean field $\mvB=\sum_i\langle a^i\rangle
\vb^i$ will also be zero,
\begin{equation}
\mvB=0\ .
\label{eq:mbzero}
\end{equation}
Ultimately, this may be seen as a consequence of the fact that both $\vB$ and
$-\vB$ are solutions of the MHD equations for given $\vV$ and since there is
always a (possibly very small) transition probability between the two states, 
a full ensemble average $\mvB$ of a statistically steady dynamo will be zero. 
This is also true if $\langle\cdot\rangle$ is a time average, provided the 
average is over a sufficiently long time, so that all dynamical states of the 
system are sampled. For the geodynamo that would mean an average over a time 
interval containing a large number of reversals (or cycle periods in case of 
the solar dynamo).

From a practical point of view, it is therefore not obvious what a time- or an
ensemble-averaged field $\mvB$ is telling us about the actual field $\vB$ of a
dynamo, and this is one of the main motivations for the approach presented in
this paper. A similar objection may be held against the mean mode coefficients
$\langle a^k\rangle$. They also do not contain much information since they are
zero (after some time), but they are indispensable for setting up the theory. 
The cross correlation coefficients $\langle a^ka^\ell\rangle$ that we compute 
in the next Section are much more useful in this regard.
\vspace{5.mm}

%%%%%%%%%%%%%%%%%%%%%%%%%%%%%%%%%%%%%%%%%%%%%%%%%%%%%%%%

\subsection{Mode excitation and phase mixing}
\label{sec:phasemix}
The interpretation of these results is the following \cite{H03}. The
eigenmodes of the dynamo equation serve as a kind of normal modes of the
dynamo. These normal modes are transiently excited and have a frequency
$\Im\lambda_k$. The relative r.m.s. mode amplitudes will be computed in the
next Section. Random phase shifts due to the stochastic forcing render the
modes quasi-periodic with a coherence time $-1/\Re\lambda_k$. Due to the
averaging, these phase shifts appear as a damping of the mean mode amplitude
$\langle a^k\rangle$. This is a familiar phenomenon in statistical physics
known as phase mixing.

%%%%%%%%%%%%%%%%%%%%%%%%%%%%%%%%%%%%%%%%%%%%%%%%%%%%%

\section{Mean square mode coefficients}
\label{sec:rmsmod}
In this Section we are interested in the cross-correlation coefficients
$\langle a^ka^{\ell*}\rangle$. The strategy is to derive an equation of the
type $(\md/\md t)\{a^ka^{\ell*}\}={\rm operator}\cdot\{a^ia^{j*}\}$. We may
then apply relation (\ref{eq:stocheq2}) to find the equation for $\langle
a^ka^{\ell*}\rangle$ (see Ref.~\cite{vK76}, \S\ 16). The analysis is
straightforward (we use $\dot{\ }=\md/\md t\,)$:
\begin{eqnarray}
(a^ka^{\ell*})\dot{\ }
&=&\dot{a}^k a^{\ell*}\,+\,a^k \dot{a}^{\ell*}
\nonumber \\[2.mm]
&=&\left(S^{ki}+C^{ki}\right)a^ia^{\ell*}\,+\,
a^k\left(S^{\ell j*}+C^{\ell j*}\right)a^{j*}
\nonumber \\[2.mm]
&=&\mbox{\LARGE(}\underbrace{S^{ki}\delta^{\ell j}\,+\,
                             S^{\ell j*}\delta^{ki}}_A
\nonumber \\
&&\qquad\quad+\,\,\underbrace{C^{ki}\delta^{\ell j}\,+\,
                             C^{\ell j*}\delta^{ki}}_{F(t)}
   \mbox{\LARGE)}\;a^ia^{j*}\ .
\label{eq:dakal}
\end{eqnarray}
This is of the form (\ref{eq:stocheq1}) with $x=\{a^ka^{\ell*}\}$. The
conditions of a short correlation time $F\tauc\ll 1$ and $A\tauc\ll 1$
are the same as (\ref{eq:ckltau}) and (\ref{eq:rkltau}), respectively. The
application of recipe (\ref{eq:stocheq2}) for the average is mainly a matter
of bookkeeping of upper indices:
\begin{widetext}
\begin{eqnarray}
\frac{\md}{\md t}\,\bigl\langle a^ka^{\ell*}\bigr\rangle&=&
\biggl\{S^{km}\delta^{\ell n}\,+\,S^{\ell n*}\delta^{km}\,+
\int_0^\infty\!\md\tau\;\Big\langle\Big(C^{ki}(t)\,\delta^{\ell j}+
C^{\ell j*}(t)\,\delta^{ki}\Big)\cdot
\nonumber \\[2.mm]
&&\qquad\qquad\qquad\qquad\qquad\qquad\qquad\cdot\,
\Big(C^{im}(t-\tau)\,\delta^{jn}+C^{jn*}(t-\tau)\,\delta^{im}
\Big)\Big\rangle\biggr\}\,\bigl\langle a^ma^{n*}\bigr\rangle
\nonumber \\[3.mm]
&=&\biggl\{S^{km}\delta^{\ell n}\,+\,S^{\ell n*}\delta^{km}\,+
\int_0^\infty\!\md\tau\;\Big(\bigl\langle C(t)\cdot C(t-\tau)
\bigr\rangle^{km}\,\delta^{\ell n}\,+\,\bigl\langle C(t)\cdot
C(t-\tau)\bigr\rangle^{\ell n*}\,\delta^{km}\Big)
\nonumber \\[2.mm]
&&\qquad\qquad\qquad\qquad\qquad +\int_0^\infty\!\md\tau\;
\Big(\bigl\langle C^{km}(t)C^{\ell n*}(t-\tau)\bigr\rangle\,+\,
\bigl\langle C^{\ell n*}(t)C^{km}(t-\tau)\bigr\rangle\Big)
\biggr\}\,\bigl\langle a^ma^{n*}\bigr\rangle \qquad
\nonumber \\[5.mm]
&=&\biggl\{D^{km}\delta^{\ell n}\,+\,D^{\ell n*}\delta^{km}\,+
\int_0^\infty\!\md\tau\;\Big(\bigl\langle C^{km}(t)C^{\ell n*}
(t-\tau)\bigr\rangle\,+\,\bigl\langle C^{\ell n*}(t)C^{km}(t-\tau)
\bigr\rangle\Big)\biggr\}\,\bigl\langle a^ma^{n*}\bigr\rangle\ .
\label{eq:dmakal1}
\end{eqnarray}
\end{widetext}
We see that the dynamo operator $D$ also plays a role in the evolution of
the mean square mode coefficients. Relation (\ref{eq:dmakal1}) becomes more
transparent if we adopt the eigenfunctions of $\na\times D$ as our basis:
\begin{eqnarray}
\biggl(\frac{\md}{\md t}\,-\,\lambda_k &-&\lambda_\ell^*\biggr)
\langle a^ka^{\ell*}\rangle\,=\qquad\qquad\qquad\qquad\qquad
\nonumber \\[1.mm]
&&\quad\left(M^{km\ell n}+M^{\ell nkm*}\right)\langle a^ma^{n*}\rangle\ .
\label{eq:dmakal2}
\end{eqnarray}
(summation over $n,m$, not over $k,\ell$). The matrix $M^{km\ell n}$ is
defined as
\begin{equation}
M^{km\ell n}\,=\,\int_0^\infty\!\md\tau\,
\langle C^{km}(t)C^{\ell n*}(t-\tau)\rangle\ .
\label{eq:mkmln}
\end{equation}

%%%%%%%%%%%%%%%%%%%%%%%%%%%%%%%%%%%%%%%%%%%%%%%

\subsection{Interpretation of Eq.~(\ref{eq:dmakal2})}
\label{sec:interpr}
Equation~(\ref{eq:dmakal2}) describes the time evolution of the mean square
mode coefficients $\langle a^ka^{\ell*}\rangle$. We replace $\md/\md t$ by
$\Lambda$, as usual, to obtain the eigenvalue problem (the double indices
$k,\ell$ and $m,n$ may each be grouped into a single new index). Of interest
is the largest eigenvalue $\Lambda$, which should be approximately zero,
otherwise $\langle a^ka^{\ell*}\rangle$ would either grow indefinitely, or
become zero. This is again a matter of using the correct values of the dynamo
coefficients. For instance, in a statistically steady geodynamo simulation the
magnetic energy is on average constant, which implies that the mean square
mode coefficients $\langle a^ka^{\ell*}\rangle$ are constant, even though some
of the modes may be quasi-periodic. A determination of the dynamo coefficients
$\alpha_{ij}$ and $\beta_{ijk}$ from a measurement of velocity correlation
functions in this dynamo must then result in Eq.~(\ref{eq:dmakal2}) having a
stationary solution. This solution is the eigenvector belonging to the
eigenvalue $\Lambda=0$ and it specifies the relative magnitudes of the
correlation coefficients $\langle a^ka^{\ell*}\rangle$ (the absolute level is
out of reach and requires explicit inclusion of nonlinear effects). These
contain a lot of information, such as the distribution of magnetic energy over
spatial scales. All other eigenvalues $\Lambda$ should have negative real
parts, and correspond to transient initial states.

%%%%%%%%%%%%%%%%%%%%%%%%%%%%%%%%%%%%%%%%%

\subsection{Computation of $M^{km\ell n}$}
\label{sec:Mkmln_anal}
Expression (\ref{eq:mkmln}) for $M^{km\ell n}$ looks very similar to the
correlation function in (\ref{eq:dotma}), but there is an important
difference: there is no internal summation over modes. The closure relation
is therefore no longer a comrade in arms, which renders evaluation of
(\ref{eq:mkmln}) much less straightforward. We insert (\ref{eq:ckl}) in
(\ref{eq:mkmln}):
\begin{eqnarray}
M^{km\ell n}&=&\int_0^\infty\!\md\tau\,
\langle C^{km}(t)C^{\ell n*}(t-\tau)\rangle
\nonumber \\[2.mm]
&=&\int\!\!\int_V\md{\it 1}\md{\it 2}\;
(\vhj^k\times\vb^m)_{p1}\,(\vhj^\ell\times\vb^n)_{q2}^*\cdot\qquad
\nonumber \\[2.mm]
&&\qquad\qquad\qquad\quad\cdot\,\int_0^\infty\!\md\tau\,
\langle u_{p1}^tu_{q2}^{t-\tau}\rangle\ .
\label{eq:corrfiem1}
\end{eqnarray}
Here ${\it 1}=\vvr_1$, ${\it 2}=\vvr_2$, index $p1$ indicates $p$-th vector
component at position $\vvr_1$, $u_{p1}^t=p$-th vector component of $\vu$ at
position $\vvr_1$ and time $t$, etc. Summation over the vector indices $p,q$
is implied.

There is an opportunity to simplify relation (\ref{eq:corrfiem1}) if the
correlation length $\lc$ of the flow $\vu$ is much smaller than the size of
the dynamo -- in other words, if there is a separation of spatial scales. In
that case we put $\vvr_2=\vvr_1+\vrho$ and note that $\langle u_{p1}^t
u_{q2}^{t-\tau}\rangle$ is zero if $\rho\geq\lc$. Since $\vhj^k\times\vb^m$
and $\vhj^\ell\times\vb^n$ are both of much larger spatial scale ($\sim$ size
dynamo), they are, from their point of view, either evaluated at virtually
the same location or the correlation function is zero. It follows that
\begin{equation}
M^{km\ell n}\,\simeq\,\int_V\md^3\vvr\;
(\vhj^k\times\vb^m)_p\,(\vhj^\ell\times\vb^n)_q^*\;\sigma_{pq}(\vvr)\ ,
\label{eq:corrfiem2}
\end{equation}
where the tensor $\sigma_{pq}$ is given by
\begin{equation}
\sigma_{pq}(\vvr)\,=\,\int_V\md^3\vrho\int_0^\infty\!\md\tau\,\langle
u_p(\vvr,t)\,u_q(\vvr+\vrho,t-\tau)\rangle\ .
\label{eq:corsig1}
\end{equation}
The integration over $\vrho$ extends over a small radius $\rho\sim\lc\ll$
size dynamo, centred on $\vvr$. Relations (\ref{eq:corrfiem2}) and
(\ref{eq:corsig1}) are a useful starting point for model dynamos with small
scale turbulence, and we refer to Appendix~\ref{sec:relneg} for an
application. But often, for example in numerical geodynamo models, there is no
separation of spatial scales. The theory developed here remains valid, but the
only option for computing $M^{km\ell n}$ seems to be through measurements of
the turbulent flow $\vu$. These permit computation of time series $C^{km}(t)$
from the definition (\ref{eq:ckl}), after which $M^{km\ell n}$ may be found
as an integral over the cross correlation function $\langle C^{km}(t)
C^{\ell n*}(t-\tau)\rangle$.

%%%%%%%%%%%%%%%%%%%%%%%%%%%%%%%%%%%%%

\begin{figure}
%\vspace{4.cm}
\centerline{\includegraphics[width=7.5cm]{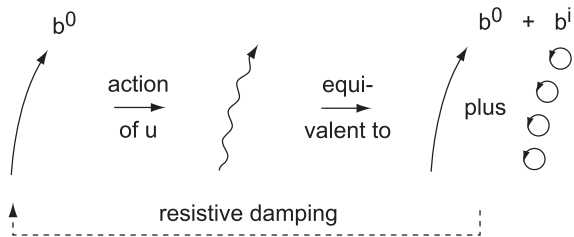}}
\caption{\small Intuitive picture of excitation of overtones in case of a
dominant fundamental mode.
\vspace{5.mm}}
\label{fig:modexc}
\end{figure}

%%%%%%%%%%%%%%%%%%%%%%%%%%%%%%%%%%%%%

\section{Mean overtone excitation level}
\label{sec:omag}
The idea of determining mean square mode coefficients $\langle a^ka^{\ell*}
\rangle$ of a statistically steady dynamo by solving the eigenvalue problem
corresponding to Eq.~(\ref{eq:dmakal2}) may be conceptually helpful, but in 
practice it leads to an extremely complicated problem because there are so 
many indices: each single index in Eq.~(\ref{eq:dmakal2}) actually comprises 
{\rm three} other indices corresponding to the three spatial degrees of 
freedom. Approximation methods seem to be the only way out. For example, if 
the geodynamo has a dominant dipole mode we may hope that the sums on the 
right hand side of (\ref{eq:dmakal2}) are dominated by the term $m=n=0$. 
Assuming a statistically steady state we then arrive at
\begin{equation}
\langle a^ka^{\ell*}\rangle\,\simeq\;-\;
\frac{M^{k0\ell0}+M^{\ell0k0*}}{\lambda_k\,+\,\lambda_\ell^*}\;
\langle\vert a^0\vert^2\rangle\ .
\label{crosscorr}
\end{equation}
This relation gives the magnitude of the cross correlation $\langle a^k
a^{\ell*} \rangle$ relative to the mean square amplitude of the fundamental
mode. The mean square overtone amplitude is:
\begin{equation}
\langle\vert a^k\vert^2\rangle\,\simeq\;
\frac{M^{k0k0}}{-\,\Re\lambda_k}\;\langle\vert a^0\vert^2\rangle\ .
\label{autocorr}
\vspace{2.mm}
\end{equation}
Since $M^{k0k0}=\int_0^\infty\!\md\tau\,\langle C^{k0}(t)C^{k0*}(t-\tau)
\rangle\simeq\vert C^{k0}\vert_{\rm r.m.s.}^2\tauc\allowbreak=\big\vert
\int_V\,\vu\cdot\vhj^k\times\vb^0\,\md^3\vvr\big\vert_{\rm r.m.s.}^2
\tauc\ $ we obtain
\vspace{2.mm}
\begin{equation}
\frac{\langle\vert a^k\vert^2\rangle}{\langle\vert a^0\vert^2\rangle}\;
\simeq\;\frac{-1}{\Re\lambda_k}\;\left\vert\int_V\,\vu\cdot\vhj^k\times
\vb^0\,\md^3\vvr\,\right\vert_{\rm r.m.s.}^2\tauc
\label{eq:aka0est1}
\vspace{2.mm}
\end{equation}
The physical picture behind (\ref{eq:aka0est1}) is as follows, see
Fig.~\ref{fig:modexc}. The turbulent convection $\vu$ creates new small scale
fields (overtones) by continuously deforming the dominant fundamental mode.
This is encoded in the numerator of (\ref{eq:aka0est1}). These overtones are
subsequently damped by resistive effects ($\Re\lambda_k$ in the denominator).
The balance of the two processes determines the excitation level. Excitation
of overtone $k$ is effective at locations where $\vhj^k\times\vb^0$ is large,
in particular when the turbulent flow $\vu$ is parallel to it. Hoyng and Van
Geffen \cite{HvG93} have derived a relation similar to (\ref{eq:aka0est1}) for
a simple dynamo, and these authors found that it agrees accurately with
numerical results.

Further progress requires evaluation of (\ref{eq:aka0est1}) or of $M^{k0k0}$,
for which there are basically two options. For a model dynamo it may be done
analytically. Otherwise we need to measure the flow $\vu$ and compute
(\ref{eq:aka0est1}) as an integral over the correlation function of
$C^{k0}(t)$.

Here we shall restrict ourselves to an estimate of the order of magnitude of
(\ref{eq:aka0est1}) for a spherical dynamo with radius $R$. The volume
integration in (\ref{eq:aka0est1}) may be estimated as in (\ref{eq:ckltau}),
except that we now do not eliminate the number of convective cells $N$:
\begin{eqnarray}
C^{k0}&\sim&\int_V\,\vu\cdot\vhj^k\times\vb^0\,\md^3\vvr\;
\nonumber \\[1.mm]
&\sim&u_{\rm r.m.s.}\,N^{-1/2}\,(R/k)^{-1}\ ,
\label{eq:estim}
\end{eqnarray}
so that $\vert\int\cdots\vert_{\rm r.m.s.}^2\tauc\sim u_{\rm r.m.s.}^2\tauc\,
N^{-1}(k/R)^2\sim (\beta/R^2)\,N^{-1}k^2$, where $\beta$ is the turbulent
diffusion coefficient. Due to the resistive term in (\ref{eq:rkl}),
$\Re\lambda_k$ scales approximately as $k^2$, so that $k^2/(-\Re\lambda_k)\sim$
constant $\sim R^2/\beta$. Hence we predict that to order of magnitude
\begin{equation}
\frac{\langle\vert a^k\vert^2\rangle}{\langle\vert a^0\vert^2\rangle}\;
\sim\;\frac{\beta}{R^2}\;\frac{1}{N}\;\frac{k^2}{-\Re\lambda_k}\,
\sim\,\frac{1}{N}\ \ \ \ {\rm for}\ \ \ \ k\ll\frac{R}{\lc}\ .
\label{eq:aka0est2}
\end{equation}
The meaning of $k$ above and in (\ref{eq:ckltau}) and (\ref{eq:rkltau}) is not
immediately clear as it comprises three quantum numbers $n$ (radial) and $\ell,
\,m$ for the two angular co-ordinates. Since $k$ appears when $R/k$ is taken to
be the spatial scale of $\vb^k$, one should assign $k=1$ to the fundamental mode.
For overtones one may think of $k$ as a geometrical mean, $k=(n\ell m)^{1/3}$,
but an exact interpretation is of course not possible.

The implication for the geodynamo is that the r.m.s. mode amplitude relative to
the fundamental mode would be of the order of $N^{-1/2}\sim 0.1$, approximately
independent of mode number. This does not seem to be an outrageous number, but
it cannot be readily compared with the data \cite{MMM96} because these do not
distinguish between modes of different radial order.

Estimate (\ref{eq:aka0est2}) assumes effectively that $k\ll R/\lc$, i.e. that
$R/k$, the spatial scale of mode $\vb^k$, is much larger than the correlation
length $\lc$. For high-order overtones with $k\gg R/\lc$ (spatial scale of mode
$\vb^k$ smaller than $\lc$) we obtain:
\begin{equation}
\frac{\langle\vert a^k\vert^2\rangle}{\langle\vert a^0\vert^2\rangle}\;\sim\;
\frac{1}{N}\;\left(\frac{R}{k\lc}\right)^3\,\qquad{\rm for}\qquad
k\gg\,\frac{R}{\lc}\ .
\label{eq:aka0est3}
\end{equation}
For very high order modes the excitation level approaches zero.

%%%%%%%%%%%%%%%%%%%%%%%%%%%%%%%%%%%%%%%%%%%%%%%%%

\section{Discussion}
\label{sec:disc}
We have analysed the statistical properties of the magnetic field generated
by a turbulent dynamo in a nonlinearly saturated state. The properties of the
mean and convective flow in this saturated state are supposed to be given,
which allows us to use linear theory. Starting from an expansion in a set of
base functions, we have derived statistical properties of the expansion
coefficients $a^k(t)$, viz. the means $\langle a^k\rangle$, the cross
correlations $\langle a^ka^{\ell*}\rangle$, and the autocorrelation functions
$\langle a^k(t)a^{k*}(t+\tau)\rangle$. The convection may have any spatial
scale distribution. Conditions for validity are (1) a short correlation time,
$C^{k\ell}\tauc\ll 1$, and (2) the effect of the mean flow $\vv$ in one
correlation time is small, $S^{k\ell}\tauc\ll 1$. The second assumption is
made for convenience, and the theory can still be deployed if it does not
hold. These two conditions have been worked out for the geodynamo in
(\ref{eq:ckltau}) and (\ref{eq:rkltau}), and seem to be well satisfied for the
lower multipole coefficients $a^k$. The two main tools enabling our analysis
are (1) the fact that currents and vector potentials form a biorthogonal set
in the volume $V$ of the dynamo, and (2) the theory of stochastic differential
equations.

Any set of magnetic fields $\vb^k(\vvr)$ may be used for the expansion.
However, since the theoretical results contain averages over the turbulence
that also occur in mean field theory, it turns out that the eigenfunctions of
the dynamo equation are a preferred set, in terms of which our results assume
their simplest form. This is an important point, and the reader might easily
get the wrong impression, as we pay quite some attention to the dynamo
coefficients $\alpha_{ij}$ and $\beta_{ijk}$ and to eigenfunctions and
eigenvalues of the dynamo equation. But this is only done in the interest of
determining the preferred basis and its properties, not because we want to
focus our study on the mean field concept $\mvB$.

%%%%%%%%%%%%%%%%%%%%%%%%%%%%%%%%%%%%%%

\subsection{Nature of the dynamo field}
\label{sec:naturefield}
The physical picture that emerges from our study is that the dynamo field is
a superposition of transiently excited eigenmodes whose coherence time and
frequency are determined by the corresponding eigenvalue of the dynamo
equation, as had been surmised by Hoyng \cite{H03}. For the theory to make
sense, all these eigenvalues should have negative real parts. This property
could be proven in the restricted case of locally isotropic turbulence, but a
general proof is still lacking. The r.m.s. excitation level of the modes is
given by an equation that determines the cross correlations $\langle a^k
a^{\ell*}\rangle$ up to an overall constant. It follows that the relative
excitation level of the modes is determined by the statistics of the flow and
by linear theory. This is an example showing that linear theory is not yet
fully exhausted. Determination of the absolute excitation levels requires
explicit inclusion of nonlinear effects in the theory. Unfortunately, this
equation for $\langle a^k a^{\ell*}\rangle$ is rather complicated and cannot
be solved in general. An approximate expression for $\langle\vert a^k\vert^2
\rangle\,/\,\langle\vert a^0\vert^2\rangle$ exists when the fundamental mode
is dominant in magnitude, which may hopefully apply in the case of numerical
geodynamo models.

%%%%%%%%%%%%%%%%%%%%%%%%%%%%%%%%%%%%%%

\subsection{Applications}
\label{sec:applic}
Application of this work is restricted to kinematically stable dynamos, as 
defined in Sec.~\ref{sec:intro}, that are in a quasi-steady saturated state.
Otherwise, there is no restriction, and numerical models and laboratory dynamo 
experiments are equally well eligible. And although current computer 
technology does not yet permit construction of numerical solar dynamo models 
with adequate resolution, such models do also qualify once they become 
available (and turn out to be kinematically stable, which in view of their 
much higher magnetic Reynolds number could be a problem). What is needed is 
the possibility to measure the flow, so that the dynamo coefficients 
$\alpha_{ij}$ and $\beta_{ijk}$ and thence the preferred basis may be 
determined. The theory developed here may be extended in several directions. 
One is the computation of the mean reversal rate of numerical geodynamo 
models, see Sec.~\ref{sec:revrate}. Another would be the statistical 
distribution of the expansion coefficients. We did not consider this topic 
here, but it might be of interest as a theoretical underpinning of the Giant 
Gaussian Process approach to geomagnetic field modelling \cite{CP88,HlM94}.

%%%%%%%%%%%%%%%%%%%%%%%%%%%%%%%%%%%%%%%%%%%%%%%%%%%%%

\subsection{Mean reversal rate and phase memory}
\label{sec:revrate}
The geomagnetic dipole reverses its direction at random moments, on average
once every few $10^5$ yr \cite{MMM96}, and numerical geodynamo models exhibit 
a similar behaviour. In this paper we have argued that the imaginary part of
$\lambda_k$ is the frequency and $-1/\Re\lambda_k$ the coherence time of mode
$k$. We apply that to the fundamental mode, and identify the coherence time of
the fundamental dipole mode with the mean time between reversals. So we
anticipate that $\lambda_0$ is real (zero frequency) and slightly negative (a
relatively long coherence time). This leads to a simple method for computing
the mean reversal rate of a geodynamo model: determine the $\alpha$ and 
$\beta$ tensors (by measuring the flow) and compute $\lambda_0$.

Unfortunately, there is a snag. In units of the inverse diffusion time
$\beta/R^2$, we will have $0<-\lambda_0\ll 1$, while for the overtones
$-\Re\lambda_k\geq 1$. It will be difficult to determine the dynamo
coefficients accurately enough to attain a precision in the eigenvalues of a 
small fraction of $\beta/R^2$. And this is necessary since the mean reversal 
rate of the geodynamo is $\sim 0.01\beta/R^2$. So this does not seem to be a 
practical way to compute these quantities. However, if the fundamental mode is 
dominant we may use (\ref{crosscorr}) with $k=\ell=0$, to obtain:
\begin{equation}
{\rm mean\ reversal\ rate}\,=\,-\,\lambda_0\,\simeq\,M^{0000}\,
\sim\,\frac{\beta}{R^2}\;\frac{1}{N}\ .
\label{eq:perrev}
\end{equation}
Here we used (\ref{eq:estim}) to estimate $M^{0000}\sim\vert C^{00}
\vert^2_{r.m.s.}\tauc\sim (\beta/R^2)/N$. For the geodynamo (\ref{eq:perrev})
is of the right order of magnitude, $\sim 8\times 10^{-14}\,$s$^{-1}$ or once
per $4\times 10^5\,$yr, taking for $\beta\sim 100\,\eta\sim 100\,$m$^2$s$^{-1}$ 
and $N\sim 100$ convection cells. We mention in passing that the determination 
of the mean reversal rate from data is complicated by issues such as the time 
resolution of the data \cite{AS94}.

A related application would be the solar dynamo, which has a periodic
fundamental mode, $\Im\lambda_0\simeq (22\,{\rm yr})^{-1}$, and according to 
the theory developed here, $-1/\Re\lambda_0$ would be the coherence time of 
that mode, i.e. the time over which the solar cycle remembers its phase. From 
observations we know that the variability of the period $P$ of the solar cycle 
$(\delta P)_{\rm r.m.s.}/P\simeq\Re\lambda_0/\Im\lambda_0\sim 0.1$ \cite{H96}. 
In principle it should now be possible to compute this number from theory.

%%%%%%%%%%%%%%%%%%%%%%%%%%%%%%%%%%%%%%%%%

\subsection{The FOSA enigma}
\label{sec:enigma}
Although mean field dynamo theory is not our main focus, we point out that we 
are now in a position to shed new light on the problem of the First Order 
Smoothing Approximation (FOSA). We illustrate the point for the solar dynamo, 
but the argument most likely holds for any turbulent dynamo. The FOSA enigma 
refers to the derivation of the dynamo equation, where one is forced to make 
an unjustified approximation, the First Order Smoothing Approximation (FOSA). 
But in spite of this the dynamo equation produces very convincing results, 
such as a periodic solar dynamo with migrating dynamo waves, a butterfly 
diagram, etc. In short, mean field theory seems to perform much better than 
one may reasonably expect, and the question is why?

A fast road to mean field theory is to interpret (\ref{eq:stocheq1}) as the
induction equation, i.e. we put $x=\vB$, $A=\na\times\vv\,\times\allowbreak
-\eta\na^2$ and $F(t)=\na\times\vu\,\times\,$, see 
Appendix~\ref{sec:appa}.\ref{sec:appaf} or Ref.~\cite{H03}. The result is the 
dynamo equation (\ref{eq:dyneq}), and since $F=(\na\times\vu\,
\times)_{\rm r.m.s.}\simeq u_{\rm r.m.s.}/\lc$, the short correlation time 
requirement $F\tauc\ll 1$ leads to $u_{\rm r.m.s.}\tauc/\lc\ll 1$ (FOSA), 
which is unlikely to be satisfied in actual dynamos.

One should be aware of the fact that the condition of a short correlation
time, $F\tauc\ll 1$, is qualitative. It is not known by how much
$F\tauc$ should be smaller than unity to avoid that 
higher-than-second-order correlations become important, and this may also 
differ from application to application. It is possible that in the dynamo 
case $F\tauc=fu_{\rm r.m.s.}\tauc/\lc$ with a numerical factor $f$ that is 
actually much less than unity. This could explain why the solutions of the 
dynamo equation behave as if $F\tauc\ll 1$ even though $u_{\rm r.m.s.}
\tauc/\lc\sim 1$. Unfortunately, it is not obvious how this idea may be 
verified.

The theory developed here does not consider an average of $\vB$, but rather
averages of expansion coefficients of $\vB$, and that makes a big difference.
This allows us to shed some light on the FOSA issue from a different
perspective. The short correlation time condition (\ref{eq:ckltau})
translated to the solar dynamo is now
\begin{eqnarray}
C^{k\ell}\tauc&\sim&u_{\rm r.m.s.}\,\tauc\left(
\frac{4\pi R^2d}{\lc^3}\right)^{-1/2}\!\!\frac{1}{\sqrt{Rd\,}/k}
\nonumber \\[2.mm]
&\sim&\frac{u_{\rm r.m.s.}\,\tauc}{\lc}\;\frac{k}{\sqrt{4\pi}}\,
\left(\frac{\lc}{R}\right)^{3/2}\frac{\lc}{d}\;\ll\,1\ .
\label{eq:cklsun}
\end{eqnarray}
For the number of convection cells we took $4\pi R^2d/\lc^3$ with $d=$
thickness of the dynamo layer, and $\sqrt{Rd\,}=$ spatial scale of the
fundamental mode, $\sqrt{Rd\,}/k$ for overtone $k$. The remarkable fact is
that (\ref{eq:cklsun}) is amply satisfied, while FOSA is not ($u_{\rm r.m.s.}
\tauc\sim\lc$). It follows that the results of the present paper hold for the
solar dynamo. The magnetic field consists therefore of a superposition of
transiently excited eigenmodes of the dynamo equation (\ref{eq:dyneq}) with
$\alpha_{ij}$ and $\beta_{ijk}$ given by (\ref{eq:albet}) {\em even though
FOSA is violated}. The FOSA problem would be fully solved if we can show that
the fundamental mode has a dominant amplitude, that is, in the stationary
solution $\langle a^k a^{\ell*}\rangle$ of Eq.~(\ref{eq:dmakal2}), the largest
element should be $\langle\vert a^0\vert^2\rangle$. We have thus eased the
FOSA problem by mapping it onto another problem that seems more amenable to a
quantitative treatment.

%%%%%%%%%%%%%%%%%%%%%%%%%%%%%%%%%%%%%%%%%

\subsection{Outlook}
\label{sec:outlook}
We are currently in the process of testing the theory developed here with the
help of a numerical geodynamo model. The averages we have computed above
depend only on a few properties of the turbulent convection: the mean flow 
$\vv$ and the dynamo coefficients $\alpha_{ij}$ and $\beta_{ijk}$. These may 
be inferred from sufficiently long measurements of the flow $\vu(t)$. This 
allows computation of $\alpha_{ij}$ and $\beta_{ijk}$ from their defining 
relation (\ref{eq:albet}), and the eigenfunctions and eigenvalues of the 
dynamo equation. Once we know the basis functions, we may obtain time series 
of $C^{k\ell}(t)$ and, finally, $M^{km\ell n}$ from (\ref{eq:mkmln}).
It seems unlikely, incidentally, that $M^{km\ell n}$ can be computed with
(\ref{eq:corrfiem2}) and (\ref{eq:corsig1}), because there appears to be no
clear separation of scales in numerical models. At the same time the dynamo
field $\vB$ is to be measured and projected onto $\vhb^k$ [or rather the
vector potential $\vA$ is to be projected on $\vhj^k$ as in (\ref{eq:akt2})]
to obtain the expansion coefficients $a^k(t)$ and their statistical
properties. These may then be compared with the theoretical predictions. The
usual practice in numerical dynamo models is that one observes and analyses
interaction between {\em local} structures. A novel aspect of our approach is
that the physics of the dynamo may now be analysed in terms of interaction
of {\em global} structures. It is hoped that in doing so we gain new
perspectives on the inner workings of numerical dynamo models.

%%%%%%%%%%%%%%%%%%%%%%%%%%%%

\section*{Acknowledgements}
I am obliged to Drs. D. Schmitt, M. Schrinner, R. Cameron and Prof. G. Barkema
for many useful discussions. The paper has benefitted considerably from the 
constructive comments of the (unknown) referees. Part of this work has been 
done while I was a guest at the Max-Planck-Institut f\"ur 
Sonnensystemforschung. I thank the MPS for its hospitality and support. 

%%%%%%%%%%%%%%%%%%%%%%%%%%%%
\appendix
%%%%%%%%%%%%%%%%%%%%%%%%%%%%

\section{The form of the mode equations}
\label{sec:formmod}
Consider first an inner product in $V$, and construct an adjoint set $\vhb^i$ 
of $\vb^j$ so that these are biorthogonal in $V$: $\int_V \vhb^i\cdot\vb^j\,\md^3\vvr
\,=\,\delta^{ij}$. The equivalent of (\ref{eq:biorth1}) is:
%
%\begin{eqnarray}
\begin{equation}
\int_V\vhb_1\cdot\vb_2\,\md^3\vvr\,=
%\nonumber \\[1.mm]
\int_V\vhj_1\cdot\va_2\,\md^3\vvr\,-
\oint_{\pa V}\vhb_1\times\va_2\cdot\md^2\vsig\ .
\label{eq:biorth1a}
\end{equation}
%\end{eqnarray}
%
The surface term is in general nonzero, and is no longer cancelled by a second 
surface term $\oint_{\pa E}...$. Hence currents and vector potentials no 
longer constitute a biorthogonal set in $V$. However, we may still proceed and 
infer from (\ref{eq:dota1}) and (\ref{eq:induct}) that
\begin{eqnarray}
\dot{a}^k(t)&=&\int_V\,\vhb^k\cdot\dot{\vB}\;\md^3\vvr
\nonumber\\[2.mm]
&=&\int_V\,\vhb^k\cdot\na\times\{(\vv+\vu)-\,\eta\na\}
\times\vB\;\md^3\vvr\ ,\qquad
\label{eq:dota3a}
\end{eqnarray}
which is (\ref{eq:dota3}). We may get rid of $\na\times$ operating on 
$\{\cdots\}$ by integrating by parts which yields (\ref{eq:dota2}), however 
at the expense of the following surface term added to the right hand side:
\[
-\oint_{\pa V}\,\vhb^k\times\{(\vv+\vu)-\,\eta\na\}
\times\vB\cdot\md^2\vsig\ ,
\]
which is in general nonzero, i.e. the boundary conditions do not force it to 
be zero. The conclusion is that if we use an inner product in $V$ there is no 
way to infer (\ref{eq:dota2}) -- we are stuck with (\ref{eq:dota3}).

Next consider an inner product in $V+E$, as defined in (\ref{eq:biorth}). This 
leads us to Eq.~(\ref{eq:dota1}), and we then hit the problem that it is not 
clear how to proceed with $\int_E\,\vhb^k\cdot\dot{\vB}\;\md^3\vvr$. The 
induction equation does not hold in $E$, and there is no obvious choice for 
$\vV=\vv+\vu$ and $\eta$ that one can make, so that $\na\times(\vV\,-\,\eta
\na\,)\times\vB$ represents $\dot{\vB}$ in $E$. Again we would be stuck, 
except that we now have the biorthogonal current - vector potential mechanism 
(\ref{eq:biorth2}) at our disposal. We integrate (\ref{eq:akt1}) by parts and 
arrive at Eq.~(\ref{eq:akt2}). There is no longer any problem with surface 
terms. Finally we proceed from  Eq.~(\ref{eq:akt2}) to (\ref{eq:dota2}) as 
outlined in the main text.

%%%%%%%%%%%%%%%%%%%%%%%%%%%%%%%%%%%%%%%%%%%%%%%%%%%%%%%%%%%%%%%%%%%

\section{Explicit form of the dynamo operator $D$}
\label{sec:appa}
There are two ways to arrive at Eqs.~(\ref{eq:dotmaf}-\ref{eq:dyneq}). The
fast way is to bypass (\ref{eq:dotma}) and to derive (\ref{eq:dyneq}) and
(\ref{eq:defd}) straightaway. We then insert expansion (\ref{eq:exp}) in
(\ref{eq:dyneq}), and use the inner product (\ref{eq:biorth2}) to isolate
Eqs.~(\ref{eq:dotmaf})-(\ref{eq:dkl}). This method a disadvantage: it is
subject to the FOSA condition and has therefore a restricted applicability.
The second method is to start from Eq.~(\ref{eq:dotma}) and to compute the
correlation function appearing in there, which leads us to (\ref{eq:dotmaf})
and following equations. This method is more laborious and also assumes a
short correlation time. However, since it uses expansion coefficients of $\vB$
rather than $\vB$ itself, it turns out to have much wider range of validity.
This validity issue is not considered here but in Sec.~\ref{sec:avmod} and
\ref{sec:enigma}.

%%%%%%%%%%%%%%%%%%%%%%%%%%%%%%%%%%%%%%%

\subsection{Computation via the dynamo equation}
\label{sec:appaf}
We interpret (\ref{eq:stocheq1}) as the induction equation, and identify
$x=\vB$, $A=\na\times\vv\,\times\allowbreak-\eta\na^2$ and $F(t)=\na\times\vu
\,\times\,$. Application of (\ref{eq:stocheq2}) leads to the dynamo equation
(\ref{eq:dyneq}) with
\begin{eqnarray}
D\vb&=&\left\{(\vv-\eta\na)\,+\int_0^\infty\!\md\tau\,
\langle\vu^t\times\na\times\vu^{t-\tau}\rangle\right\}\times\vb\qquad
\nonumber \\[2.mm]
&=&(\vv-\eta\na)\times\vb\,+\int_0^\infty\!\md\tau\,
\nonumber \\[2.mm]
&&\ \ \ \left\langle\,\vu^t\times[\,(\vb\cdot\na)\vu^{t-\tau}\,-\,
(\vu^{t-\tau}\cdot\na)\vb\,]\,\right\rangle\ ,
\label{eq:A1}
\end{eqnarray}
since $\na\cdot\vb=0$, and we assume incompressibility, $\na\cdot\vu=0$. Time
arguments appear momentarily as an upper index: $\vu^t\equiv\vu(\vvr,t)$.
Relation (\ref{eq:A1}) is equivalent to (\ref{eq:defd}) with
\begin{eqnarray}
\alpha_{ns}&=&\epsilon_{npq}\int_0^\infty\!\md\tau\,
\langle u_p^t(\na_s u_q^{t-\tau})\rangle\ ,\qquad\qquad\quad
\nonumber \\[0.mm]
\lefteqn{\rm and}
\hspace{2.cm}
\label{eq:albet}
\\[0.mm]
\beta_{nqs}&=&\epsilon_{nps}\int_0^\infty\!\md\tau\,
\langle u_p^tu_q^{t-\tau}\rangle\ ,
\nonumber
\end{eqnarray}
where $u_p^t=p$-th vector component of $\vu(\vvr,t)$. These are the familiar
$\alpha$- and $\beta$-tensors from mean field theory, in the FOSA
approximation, for incompressible convection, and without assuming any flow
symmetry. Note that $\beta_{nqs}=0$ when $n=s$. Hence 9 out of the 27
components of $\beta_{nqs}$ are zero. This is no longer the case when
resistive effects are taken into account.

To make the connection with Eq.~(\ref{eq:dotmaf}) insert expansion
(\ref{eq:exp}) in (\ref{eq:dyneq}): $\pa_t\langle a^i\rangle\vb^i=\langle
a^\ell\rangle\na\times D\vb^\ell$ (summation over $i$ and $\ell$). Uncurling
to $\pa_t\langle a^i\rangle\va^i=\langle a^\ell\rangle D\vb^\ell+\na\psi$,
left multiplication with $\vhj^k$ and use of the inner product
(\ref{eq:biorth2}) produces (\ref{eq:dotmaf}) and (\ref{eq:dkl}).

%%%%%%%%%%%%%%%%%%%%%%%%%%%%%%%%%%%%%%%%

\subsection{Computation from (\ref{eq:dotma})}
\label{sec:appae}
The disadvantage of the above argument is that the requirement of a short
correlation time $F\tauc\ll 1$ leads to the restriction $u_{\rm r.m.s.}
\tauc/\lc\ll 1$ (FOSA), while there are, unfortunately, several indications
that $u_{\rm r.m.s.}\tauc/\lc\sim 1$. We can do better if we start from
Eq.~(\ref{eq:dotma}) and compute the correlation function. The result is the
same, but the range of validity is much wider, see Sec.~\ref{sec:avmod} and 
\ref{sec:enigma}.

To evaluate the correlation function in (\ref{eq:dotma}) it is useful to set
up a formal approach, which is really an overkill here, but not in the next
Section as we deal with resistive effects. We consider operators in vector 
space with a left and a right vector gate, for example $\OC=\vu\,\times\,$,
and $\va\cdot\OC\cdot\vb=\va\cdot\vu\times\vb$. Since $(\OC\cdot\vb)_i=
(\vu\times\vb)_i=\epsilon_{isk}u_sb_k\,$ and $(\OC\cdot\vb)_i=C_{ik}b_k$ we
have $C_{ik}=\epsilon_{isk}u_s$. But $\OC$ has also a representation in
function space, denoted with upper indices. This is $C^{ik}(t)\,=\,\int_V\,
\vhj^i\cdot\vu\times\vb^k\,\md^3\vvr$, defined in (\ref{eq:ckl}). While
$C^{ik}$ is an $\infty\times\infty$ array of functions of time, $C_{ik}$ is a
$3\times 3$ array of functions of $\vvr$ and $t$.

For the computation of the infinite internal summation over $i$ in
(\ref{eq:dotma}), we consider a slightly more general question: given two
operators $P^{k\ell}$ and $Q^{k\ell}$ in function space representation,
compute $(PQ)^{ik}=P^{i\ell}Q^{\ell k}$ (summation over $\ell$). Here the
completeness relation (\ref{eq:complet3}) figures as the essential tool:
\begin{eqnarray}
(PQ)^{ik}&=&
\int_V\;\vhj^i\cdot\OP\cdot\vb^\ell\;\md^3\vvr
\int_V\;\vhj^\ell\cdot\OQ\cdot\vb^k\;\md^3\vvr
\nonumber \\[2.mm]
&=&\int\!\!\!\int_V\md{\it 1}\md{\it 2}\ \hj^i_{n1}\,P_{nm1}\,b^\ell_{m1}\,
\hj^\ell_{p2}\,Q_{pr2}\,b^k_{r2}\qquad
\nonumber \\[2.mm]
&=&\int_V\md^3\vvr\ \hj^i_n\,P_{nm}\,\epsilon_{pms}\,\na_s\,Q_{pr}\,b^k_r
\nonumber \\[2.mm]
&=&\int_V\md^3\vvr\ \hj^i_n\,P_{nm}\,(\na\times\OQ\cdot\vb^k)_m
\nonumber \\[2.mm]
&=&\int_V\md^3\vvr\ \vhj^i\cdot(\OP\cdot\na\times\OQ)\cdot\vb^k\ .
\label{eq:PQik}
\end{eqnarray}
Notation: $\hj^i_{n1}=$ $n$-th vector component of $\vhj^i(\vvr_1)$, etc. In
the second line we insert the completeness relation $\hj^\ell_{p2}\,
b^\ell_{m1}=\epsilon_{pms}\delta({\it 1}-{\it 2})\na_{s2}$, after which the
volume integral over ${\it 2}\equiv\vvr_2$ may be done with the help of the
delta-function. In the third line $\na_s$ operates on $Q_{pr}$ and $b^k_r$ as
these had originally the argument $\vvr_2$. We see that $\na\times$ appears at
the location of the sum over $\ell$. Relation (\ref{eq:PQik}) tells us that
the sum over $\ell$ in the top line is equal to an integral of an ordinary 
vector expression. It may be generalised to several internal summations, e.g.
\begin{equation}
(PQT)^{ik}\,=\,\int_V\md^3\vvr\ \vhj^i\cdot\OP\cdot\na\times\OQ\cdot
\na\times\OT\cdot\vb^k\ ,
\label{eq:PQTik}
\end{equation}
where each $\na$ operates on everything to its right.

The summation over $i$ in (\ref{eq:dotma}) is now almost trivial. With the
help of (\ref{eq:PQik}) and $\OC^t=\vu^t\,\times$ we find that
\begin{eqnarray}
C^{ki}(t)C^{i\ell}(t-\tau)\!&=&\!\int_V\md^3\vvr\ \vhj^k\cdot\OC(t)\cdot
\na\times\OC(t-\tau)\cdot\vb^\ell
\nonumber \\[1.mm]
&&\!\!\!\!\!\!\!\!\!\!\!\!\!\!\!\!\!\!\!\!=\int_V\md^3\vvr\
\vhj^k\cdot\vu^t\times\na\times\vu^{t-\tau}\!\times\vb^\ell\ .
\label{eq:ckicil1}
\end{eqnarray}
Now take $\int_0^\infty\md\tau\,\langle\ \rangle$ of (\ref{eq:ckicil1}),
combine that with $S^{k\ell}$ from (\ref{eq:rkl}), and we have proven that
the operator between $\{\ \}$ in (\ref{eq:dotma}) is equal to $D^{k\ell}=
\int_V\md^3\vvr\;\vhj^k\cdot D\cdot\vb^\ell$ with $D$ given by (\ref{eq:A1}).

%%%%%%%%%%%%%%%%%%%%%%%%%%%%%%%%%%%%%%%%

\subsection{Resistive effects}
\label{sec:appares}
Sofar we have ignored the exponential operators $\exp(\pm A\tau)$ in
(\ref{eq:stocheq2}) and (\ref{eq:dotma}) which is tantamount to ignoring all
mean flow and resistive effects on the $\alpha$ and $\beta$ tensors. Here we
investigate the influence of resistivity, with the help of the operator
technique introduced above. Instead of (\ref{eq:dotma}) our starting point is
now
\begin{eqnarray}
\frac{\md}{\md t}\,\langle a^k\rangle&=&\biggl\{S^{k\ell}\,+\,
\int_0^\infty\!\md\tau\,\big\langle C^{km}(t)\,[\,\exp(S\tau)\,]^{mn}\qquad
\nonumber \\[2.mm]
&&\qquad C^{np}(t-\tau)\,[\,\exp(-S\tau)\,]^{p\ell}\,
\big\rangle\biggr\}\,
\langle a^\ell\rangle\ ,
\label{eq:dotmares}
\end{eqnarray}
with $S^{k\ell}$ and $C^{mn}(t)$ given by (\ref{eq:rkl}) and (\ref{eq:ckl}).
There are three internal summations in (\ref{eq:dotmares}), over $m,n$ and
$p$, but as we shall see the exponential operators contain additional
summations. As a first step we compute
\begin{eqnarray}
\lefteqn{[\,P\exp(S\tau)Q\,]^{ik}\,=}
\nonumber \\[2.mm]
&&\quad=\left[\,P\,(\,1+\bullet\,S\tau+\hlf\bullet\,S\bullet S\,\tau^2+\cdots)
\bullet\,Q\,\right]^{ik}\qquad
\nonumber \\[2.mm]
&&\quad=\int_V\md^3\vvr\ \vhj^i\cdot\OP\cdot\exp(\na\times\OS\,\tau)
\cdot\na\times\OQ\cdot\vb^k\ .
\label{eq:pexprq}
\end{eqnarray}
In the first and second line $P,Q$ and $S$ are $\infty\times\infty$ arrays
with numbers as entries, and a bullet $\bullet$ indicates an internal
summation over a double upper index. To arrive at the last line we apply
relation (\ref{eq:PQTik}). As explained in the previous section, $\na\times$
appears at the location of each internal summation; $\OP,\,\OQ$ and $\OS$ are
vector operators, e.g. $\OS=(\vv-\eta\na)\times$, and a center dot indicates
an internal summation over a double lower (vector) index, i.e. the usual
vector product. Since relation (\ref{eq:pexprq}) is easily generalised, we are
now in a position to compute:
\begin{eqnarray}
\lefteqn{\int_0^\infty\!\md\tau\left[\,\bigl\langle C(t)\,
\exp(S\tau)\,C(t-\tau)\bigr\rangle\exp(-S\tau)\,\right]^{k\ell}}
\nonumber  \\[2.mm]
&&=\int_0^\infty\!\!\md\tau\int_V\md^3\vvr\;
\vhj^k\cdot\big\langle\OC^t\cdot\me^{\na\times\OS{\displaystyle\tau}}\cdot
\na\times\OC^{t-\tau}\big\rangle
\nonumber \\[1.mm]
&&\hspace{4.cm}\times\,\me^{-\na\times\OS{\displaystyle\tau}}\cdot\vb^\ell
\nonumber \\[3.mm]
&&=\int_V\md^3\vvr\;\vhj^k\cdot\bigg[\int_0^\infty\!\!\md\tau\;
\big\langle\vu^t\times\me^{\tau\eta\na^2}\cdot\na\times\vu^{t-\tau}
\big\rangle\quad
\nonumber \\[1.mm]
&&\hspace{4.cm}\times\,\me^{-\tau\eta\na^2}\bigg]\cdot\vb^\ell\ .
\label{eq:resop1}
\end{eqnarray}
The time argument appears again as an upper index; at the second $=$ sign (1) 
the integration order is reversed, (2) we use $\OC=\vu\,\times$, and (3) we 
stipulate zero mean flow so that $\na\times\OS=-\eta\na\times\na\times=
\eta\na^2$, since $\na\times\OS$ operates exclusively on vectors having zero 
divergence. Next, we combine (\ref{eq:resop1}) with $S^{k\ell}$ from 
(\ref{eq:rkl}) to see that Eq.~(\ref{eq:dotmares}) is equivalent to 
Eqs.~(\ref{eq:dotmaf}) and (\ref{eq:dkl}) with
\begin{eqnarray}
\lefteqn{D=(\vv-\eta\na)\,\times}
\nonumber \hspace{5.mm} \\[2.mm]
&&+\int_0^\infty\!\md\tau\;\big\langle\vu^t\times
\me^{\tau\eta\na^2}\cdot\na\times\vu^{t-\tau}\big\rangle\times\,
\me^{-\tau\eta\na^2}\,.\qquad
\label{eq:resop2}
\end{eqnarray}
The difference with (\ref{eq:A1}) is in the exponential operators $\exp
(\pm\tau\eta\na^2)$ that embody the effect of a finite resistivity on the
$\alpha$- and $\beta$ tensors. They are evolution operators: $\exp(\tau\eta
\na^2)\vb\equiv\vB(\vvr,\tau)$, where $\vB(\vvr,t)$ is the solution of $\pa_t
\vB=\eta\na^2\vB$ with initial condition $\vB(\vvr,0)=\vb(\vvr)$. This is
usually written in terms of the more familiar Green function formalism
\cite{KR80}. The operator technique has the advantage of being very compact.
Explicit expressions for the $\alpha$ and $\beta$ tensors may be extracted
from (\ref{eq:resop2}) with the help of a spatial Fourier transformation of
the flow $\vu$. For this and related matters we refer to \cite{H03}.

%%%%%%%%%%%%%%%%%%%%%%%%%%%%%%%%%%%%%%%%%%%%%%%%

\section{Proof of $\Re\lambda_k<0$}
\label{sec:relneg}
We close with a proof that the eigenvalues of the dynamo equation of a
statistically steady dynamo with locally isotropic small-scale turbulence all
have negative real parts. We begin with Eq.~(\ref{eq:dmakal2}), take $k=\ell$
and put $\pa_t\langle a^ka^{\ell*}\rangle=0$ on account of statistical
steadiness:
\begin{equation}
\Re\lambda_k\,\langle\vert a^k\vert^2\rangle\,=\,-\,\hlf\,
(M^{kmkn}+M^{knkm*})\langle a^ma^{n*}\rangle\ .
\label{eq:lkneg1}
\end{equation}
(summation over $m$ and $n$, not over $k$). It is essential that we restrict
attention to kinematically stable dynamos. In that case the induction equation
(on which our results are based) will generate the quasi-steady saturated
field of the dynamo, and $\langle a^ka^{\ell*}\rangle$ will actually be
constant.

Small-scale turbulence allows us to invoke (\ref{eq:corrfiem2}), and locally
isotropic convection is understood to imply that the tensor $\sigma_{pq}$ is
diagonal:
\begin{equation}
\sigma_{pq}(\vvr)\,=\,\beta\lc^3\,\delta_{pq}\ ,
\label{eq:lkneg2}
\end{equation}
with
\begin{equation}
\beta\lc^3\,=\,\frac{1}{3}\int_V\md^3\vrho\int_0^\infty\!\md\tau\,
\langle\vu(\vvr,t)\cdot\vu(\vvr+\vrho,t-\tau)\rangle\ .
\label{eq:lkneg3}
\end{equation}
The expression on the right is written as $\beta\lc^3$ because it has the
dimension of a turbulent diffusion coefficient times a correlation volume. We
use $\beta\lc^3$ as a generic symbol defined by (\ref{eq:lkneg3}). It may be
a function of position. In this approximation we have
\begin{equation}
M^{kmkn}\,=\,\int_V\md^3\vvr\;\beta\lc^3\;
(\vhj^k\times\vb^m)\cdot(\vhj^k\times\vb^n)^*\ .
\label{eq:lkneg4}
\end{equation}
Since we now also have that $M^{knkm*}=M^{kmkn}$, it follows that
\begin{eqnarray}
\Re\lambda_k\,\langle\vert a^k\vert^2\rangle&=&-\int_V\md^3\vvr\;
\beta\lc^3\;(\vhj^k\times\vb^m)\cdot(\vhj^k\times\vb^n)^*\cdot
\nonumber \\
&&\qquad\qquad\qquad\qquad\qquad\quad\cdot\,\langle a^ma^{n*}\rangle
\nonumber \\
&=&-\int_V\md^3\vvr\;\beta\lc^3\;\langle\vert\vhj^k\times\vB\vert^2\rangle\ .
\label{eq:lkneg5}
\end{eqnarray}
\vspace{0.mm}
In the last line the summation over $m$ and $n$ has been performed with the
help of relation (\ref{eq:exp}). It follows that $\Re\lambda_k<0$, provided
$\beta\lc^3$ is positive. This will be the case for almost all turbulent flows
$\vu(\vvr,t)$. We stress that the proof allows for a finite resistivity. We
have tried to extend the proof to more general dynamos, and while we have not
been unsuccesful yet, we are confident that such an extension should be
possible.

The physics behind the above proof is not some kind of stability analysis, but
rather ordinary phase mixing, cf. Sec.~\ref{sec:phasemix}. The field of the
dynamo may be represented as a sum of oscillators, $\vB=\sum_ia^i\vb^i$. The
mode amplitudes $a^i=A^i\cos(\omega^it+\varphi^i)$ have a variable magnitude
$A^i$ and a randomly drifting phase $\varphi^i$. The average over an ensemble
of dynamos $\mvB=\sum_i\langle A^i\cos(\omega^it+\varphi^i)\rangle\vb^i$ will
be zero. It follows that the mean field $\mvB$ over an ensemble of
statistically steady dynamos is zero, as in (\ref{eq:mbzero}). This in turn
implies $\Re\lambda_k<0$. Another pertinent remark in this connection is that
there are several examples of randomly perturbed or driven oscillators whose
mean amplitude is zero if the mean energy is constant \cite{vK76,H03}.

%%%%%%%%%%%%%%%%%%%%%%%%%%%
\np
%%%%%%%%%%%%%%%%%%%%%%%%%%%

%%%%%%%%%%%%%%%%%%%%%%%%%%%%%%%%%

\begin{references}

\bibitem{RH04} G.~R\"udiger and R.~Hollerbach, {\em The Magnetic Universe}
(Wiley-VCH, Weinheim, 2004).

\bibitem{GR95} G.A.~Glatzmaier and P.H.~Roberts, P.H., Nature {\bf 377},
203 (1995).

\bibitem{KB97} W.~Kuang and J.~Bloxham, Nature {\bf 389}, 371 (1997).

\bibitem{COG98} U.R.~Christensen, P.~Olson and G.A.~Glatzmaier, Geophys.
Res. Lett. {\bf 25}, 1565 (1998).

\bibitem{TMH05} F.~Takahashi, M.~Matsushima and Y.~Honkura, Science
{\bf 309}, 459 (2005).

\bibitem{CW07} U.R.~Christensen and J.~Wicht, in {\em Treatise on
Geophysics, Vol. 8}, edited by P. Olson (Elsevier, Amsterdam, 2007), p. 245.

\bibitem{KS97} A.~Kageyama and T.~Sato, Phys. Rev. E. {\bf 55}, 4617 (1997).

\bibitem{OCG99} P.~Olson, U.~Christensen and G.A.~Glatzmaier, J. Geophys.
Res. {\bf 104}, 10383 (1999).

\bibitem{KR80} F.~Krause and K.H.~R\"adler, {\em Mean Field
Magnetohydrodynamics and Dynamo Theory} (Akademie-Verlag, Berlin, 1980).

\bibitem{WO04} J.~Wicht and P.~Olson, Geochem. Geophys. Geosys. {\bf 5},
Q03H10 (2004).

\bibitem{SRSRC07} M.~Schrinner, K.H.~R\"adler, D.~Schmitt, M.~Rheinhardt and
U.R.~Christensen, Geophys. Astrophys. Fluid Dynamics {\bf 101}, 81 (2007).

\bibitem{WSH08} J.~Wicht, S.~Stellmach and H.~Harder, {\em Numerical Models
of the Geodynamo}, in {\em Geomagnetic Field Variations}, edited by K.H.
Glassmeier, H. Soffel and J. Negendank (Springer, 2008), p.~107.

\bibitem{M78} K.H.~Moffatt, {\em Magnetic Field Generation in Electrically
Conducting Fluids} (Cambridge U.P., 1978).

\bibitem{E46} W.M.~Elsasser, Phys. Rev. {\bf 69}, 106 (1946); Phys. Rev.
{\bf 70}, 202 (1946).

\bibitem{vK76} N.G.~van Kampen, Physics Reports {\bf 24C}, 171 (1976).

\bibitem{CT09} F.~Cattaneo and S.M.~Tobias, J. Fluid Mech. {\bf 621}, 
205 (2009).

\bibitem{TB08} A.~Tilgner and A.~Brandenburg, Mon. Not. R. Astron. Soc. {\bf
391}, 1477 (2008).

\bibitem{HOS01} P.~Hoyng, M.A.J.H.~Ossendrijver and D.~Schmitt, Geophys.
Astrophys. Fluid Dynamics {\bf 94}, 263 (2001).

\bibitem{H88} P.~Hoyng, Astrophys. J. {\bf 332}, 857 (1988).

\bibitem{PFL76} A.~Pouquet, U.~Frisch and J.~L\'eorat, J. Fluid Mech.
{\bf 77}, 321 (1976).

\bibitem{HS95} P.~Hoyng and N.A.J.~Schutgens, Astron. Astrophys. {\bf 293},
777 (1995).

\bibitem{MF53} P.M.~Morse, and H.~Feshbach, {\em Methods of Theoretical
Physics} (McGraw-Hill, New York, 1953).

\bibitem{SJSH08} M.~Schrinner, J.~Jiang, D.~Schmitt and P.~Hoyng, Mon. Not. R.
Astron. Soc., to be submitted.

\bibitem{D78} P.A.M.~Dirac, {\em The Principles of Quantum Mechanics}
(Oxford U.P., 1978).

\bibitem{vK74} N.G.~van Kampen, Physica {\bf 74}, 215 (1974); ibidem {\bf 74}, 
239 (1974).

\bibitem{H03} P.~Hoyng, in {\em Advances in Nonlinear Dynamos}, edited by A.
Ferriz-Mas and M. N\'u\~nez (Taylor \& Francis, New York, 2003), p. 1.

\bibitem{lowind} We use a lower index on $\lambda_i$ to avoid a summation
over $i$. Summation is implied over double upper indices, but not over mixed
upper and lower double indices.

\bibitem{H87} P.~Hoyng, Astron. Astrophys. {\bf 171}, 357 (1987).

\bibitem{vGH89} J.H.G.M.~van Geffen and P.~Hoyng, Astron. Astrophys.
{\bf 213}, 429 (1989).

\bibitem{HvG93} P.~Hoyng and J.H.G.M.~van Geffen, Geophys. Astrophys.
Fluid Dynamics {\bf 68}, 203 (1993).

\bibitem{MMM96} R.T.~Merrill, M.W.~McElhinny and P.L.~McFadden, {\em The
Magnetic Field of the Earth} (Academic Press, New York, 1996).

\bibitem{CP88} C.G.~Constable and R.L.~Parker, J. Geophys. Res. {\bf 93},
11569 (1988).

\bibitem{HlM94} G.~Hulot and J.L.~Le Mou\"el, Phys. Earth Planet. Int.
{\bf 82}, 167 (1994).

\bibitem{AS94} A.~Anufriev and D.~Sokoloff, Geophys. Astrophys. Fluid 
Dynamics {\bf 74}, 207 (1994).  

\bibitem{H96} P.~Hoyng, Solar Phys. {\bf 169}, 253 (1996).

\end{references}
\end{document}